\documentstyle[sprocl]{article}
\input epsf.tex
\def\DESepsf(#1 width #2){ \epsfxsize = #2 \epsfbox{#1}}
\def\be{\begin{equation}}
\def\ee{\end{equation}}
\def\bea{\begin{eqnarray}}
\def\eea{\end{eqnarray}}

\begin{document}
\title{ 
$B \to \eta' X_s$ in the Standard 
Model~\footnote{Talk presented by Xiao-Gang He
at the Workshop on CP Violation, Adelaide, Australia, 3-6 July, 1998}}

\vfill
\author{Xiao-Gang He$^{1,2}$ and Guey-Lin Lin$^3$}
\address{$^1$
\rm  Department of Physics, National Taiwan University,
Taipei, 10764, Taiwan
}
\address{$^2$
\rm School of Physics, University of Melbourne, Parkville, Vic. 3052,
Australia
}
\address{$^3$
\rm Institute of Physics, National Chiao-Tung University,
Hsinchu, 300, Taiwan
}
%
\maketitle\abstracts{
We study $B \to \eta' X_s$ within the framework of the Standard  
Model.
Several mechanisms
such as $b \to \eta' s g$ from the QCD anomaly, and 
$b \to \eta's $ and $B\to \eta' s \bar q$ from four-quark  
operators
are treated simultaneously.
We find that 
the first mechanism give a significant contribution to the 
branching ratio 
for $B\to \eta' X_s$, while the other 
two mechanisms account for about 15\% of the experimental value. 
The Standard Model
prediction for $B\to \eta' X_s$ is consistent with the CLEO data.    
}

The recent observation of $B\to \eta' K$~\cite{CLEO1} and 
$B\to \eta^{\prime}X_s$~\cite{CLEO2} decays with high momentum
$\eta^{\prime}$ has stimulated many theoretical 
activities~\cite{AS,HT,he1,KP,FR,excl,excl1,HZ}. 
One of the mechanisms proposed to account for this decay is
$b\to sg^*\to
sg\eta^{\prime}$~\cite{AS,HT} where the $\eta^{\prime}$ meson is  
produced 
via the anomalous 
$\eta'-g-g$ coupling. According to a previous analysis~\cite{HT},
this mechanism within the Standard Model(SM) can only account for 1/3 
of the 
measured branching ratio~\cite{CLEO2}: $B(B\to \eta^{\prime}X_s) =
(62\pm 16\pm 13)\times 10^{-5}$. 
There are also other calculations of $B \to \eta' X_s$ 
based on four-quark operators of the effective  
weak-Hamiltonian~\cite{he1,KP}. 
These contributions to the branching ratio, typically $10^{-4}$, 
are also too small to account for $B\to \eta' X_s$, 
although the four-quark-operator
contribution is capable of explaining the branching ratio for
the exclusive $B \to \eta' K$ decays~\cite{excl,excl1}.
These results
have inspired proposals for an enhanced $b\to sg$ and other 
mechanisms arising from physics 
beyond the Standard Model~\cite{HT,KP,FR}. 
In the following we report our recent analysis
using next-to-leading effective Hamiltonian and consider several
mechanisms simultaneously~\cite{hg3}. We conclude that the standard model 
 is consistent with 
experimental data from CLEO.

The quark level effective Hamiltonian for the $B\to \eta'  
X_s$ decay is 
given by~\cite{REVIEW}:
\begin{eqnarray}
H_{eff}(\Delta B=1)&=&{G_F\over  
\sqrt{2}}[\sum_{f=u,c}V_{fb}V_{fs}^*(C_1(\mu)O_1^f(\mu)+C_2(\mu)O_2^f(\mu))
\nonumber\\
&-&V^*_{ts}V_{tb}\sum_{i=3}^{6}(C_i(\mu)O_i(\mu)
+C_8(\mu)O_8(\mu))],
\label{HAMI}
\end{eqnarray}
The operators are defined in Ref.[13,14].
For numerical analyses, we use the scheme-independent 
Wilson coefficients obtained in Ref.[14].
For $m_t = 175$ GeV, $\alpha_s (m_Z^2) = 0.118$ and $\mu = m_b = 5$  
GeV,
we have
\begin{eqnarray}
&&C_1 = -0.313, \;\;C_2 = 1.150,\;\;C_3 = 0.017,\nonumber\\
&&C_4 = -0.037,\;\;
C_5= 0.010,\;\;C_6 = -0.045,\;\; 
\label{WILSON}
\end{eqnarray}
When the one-loop corrections to the
matrix elements are taken into account, the coefficients are modified to 
$C_i(\mu)+\bar{C}_i(q^2,\mu)$ with
\begin{equation}
\bar{C}_4(q^2,\mu) = \bar{C}_6(q^2,\mu) = -3\bar{C}_3(q^2,\mu)=
-3\bar{C}_5(q^2,\mu)=-P_s(q^2,\mu),
\end{equation}
where
\begin{equation}
P_s(q^2,\mu) = {\alpha_s \over 8\pi} C_2(\mu) \left ( {10\over 9} + 
G(m_c^2,q^2,\mu)\right ),
\end{equation}
with
\begin{equation}
G(m_c^2,q^2,\mu)=
4\int x(1-x) \log \left (
{m_c^2 - x(1-x)q^2\over \mu^2}\right )dx.
\end{equation}
The coefficient $C_8$ is equal to $-0.144$ at $\mu=5 \ {\rm GeV}
$~\cite{REVIEW}, and $m_c$ is taken to be $1.4$ GeV. 

Let us first work out  
the four-quark-operator contribution to 
$B\to \eta' X_s$.
We follow the approach of~Ref.[5,15] which uses factorization 
approximation to estimate various hadronic matrix elements. 
The four-quark operators
can induce three types of processes represented by
1) $<\eta'|\bar q \Gamma_1 b|B> <X_s|\bar s \Gamma_1'q|0>$,
2) $<\eta'|\bar q \Gamma_2 q|0><X_s|\bar s \Gamma b| B>$, and 
3) $<\eta' X_s|\bar s \Gamma_3q|0><0|\bar q\Gamma_3'|B>$. 
Here $\Gamma^{(')}_i$ denotes appropriate gamma matirces. 
The contribution from 1) gives a ``three-body'' type of decay,
$B\to \eta' s \bar q$. The contribution from 2) gives a ``two-body''
type of decay $b\to s\eta'$.
And the contribution from 3) is the annihilation
type which is relatively suppressed and will be neglected.
Several decay constants and form factors needed in the calculations  
are
listed below:
\begin{eqnarray}
&&<0|\bar u\gamma_\mu \gamma_5 u|\eta'> = 
<0|\bar d \gamma_\mu \gamma_5 d|\eta'>
=if_{\eta'}^u p^{\eta'}_\mu\nonumber\\
&&<0|\bar s \gamma_\mu \gamma_5 s|\eta'> = 
if_{\eta'}^s p^{\eta'}_\mu,\;\;
<0|\bar s \gamma_5 s|\eta'> = i(f_{\eta'}^u-f_{\eta'}^s) 
{m^2_{\eta'}\over 2m_s},\nonumber\\
&&f_{\eta'}^u = {1\over \sqrt{3}
} (f_1 \cos\theta_1 + 
{1\over \sqrt{2}} f_8 \sin\theta_8),\;\;
f_{\eta'}^s = {1\over \sqrt{3}}
(f_1\cos\theta_1 - \sqrt{2} f_8 \sin\theta_8),\nonumber\\
&&<\eta'|\bar q\gamma_\mu b|B>
= F_1^{Bq}(p^B_\mu + p^{\eta'}_\mu) +
(F_0^{Bq}-F_1^{Bq}) {mB^2-m_{\eta'}^2\over 
q^2} q_\mu,\nonumber\\
&&F_{1,0}^{Bq}={1\over \sqrt{3}} ({1\over \sqrt{2}} 
\sin\theta F^{B\eta_8}_{1,0}
+\cos\theta F^{B\eta_1}_{1,0}).
\end{eqnarray}
For the $\eta'-\eta$ mixing associated with decay constants above,
we have used the two-angle 
-parametrization.
The numerical values of various parameters are obtained from 
Ref. [16] with 
$f_1= 157$ MeV, $f_8=168$ MeV,
and the mixing angles $\theta_1 = -9.1^0$, $\theta_8=-22.1^0$. 
For the mixing angle associated with form factors, we used the
one-angle parametrization with\cite{FK} $\theta = - 15.4^o$,
since these form factors were calculated in that 
formulation~\cite{he1,he2}. 
In the latter discussion of $b\to sg\eta'$,  
we shall use the same parametrization in order to compare our results with  
those of earlier works
\cite{AS,HT}.  
For form factors, we assume
that $F^{B\eta_1} = F^{B\eta_8} = F^{B\pi}$ with 
dipole and monopole $q^2$ dependence for $F_1$ and $F_0$,  
respectively.
We used the running mass $m_s \approx 120$ MeV
at $\mu = 2.5$ GeV and $F^{B\pi} = 0.33$ following Ref.[9].

Using $V_{ts} = 0.038$, $\gamma = 64^0$ and $\mu = 5$ GeV, we find  
that
the branching ratios in the signal region $p_{\eta'} > 2.2$ GeV 
($m_X < 2.35$ GeV) are given by
\begin{eqnarray}
B(b\to \eta' s) = 0.9\times 10^{-4},\;\;
B(B\to \eta' s \bar q) = 0.1\times 10^{-4} 
\end{eqnarray}
The branching ratio can reach $2\times 10^{-4}$ if all parameters  
take 
values in favour of $B\to \eta' X_s$.
Clearly the mechanism by four-quark operator is not sufficient  
to explain the observed $B\to \eta^{\prime}X_s$ 
branching ratio.

We now turn to 
$b\to \eta' s g$ through the QCD anomaly. To see how the
effective Hamiltonian in Eq. (\ref{HAMI}) can be applied to calculate
this process,  
we rearrange the effective Hamiltonian such that
\begin{eqnarray}
&&\sum_{i=3}^6C_iO_i=(C_3+{C_4\over N_c})O_3
+(C_5+{C_6\over N_c})O_5\nonumber\\
&&
- 2(C_4-C_6)O_A+2(C_4+C_6)O_V+C_8O_8,
\label{GLUE}
\end{eqnarray}
where 
\begin{equation}
O_A=\bar{s}\gamma_{\mu}(1-\gamma_5)T^a b \sum_{q}\bar{q}\gamma^{\mu}
\gamma_5T^a q,
\;\;O_V=\bar{s}\gamma_{\mu}(1-\gamma_5)T^a b \sum_{q}
\bar{q}\gamma^{\mu}
T^a q.
\end{equation} 
Since the light-quark bilinear in $O_V$ carries the quantum number
of a gluon, one expects~\cite{AS} $O_V$ give contribution to
the $b\to sg^*$ form factors. In fact, by applying the QCD equation  
of motion
: $D_{\nu}G^{\mu\nu}_a=g_s\sum \bar{q}\gamma^{\mu}T^a q$,
we have $O_V=(1/ g_s)\bar{s}\gamma_{\mu}(1-\gamma_5)T^a b D_{\nu}
G^{\mu\nu}_a$. Let us write  
the effective $b\to sg^*$ vertex as
\begin{eqnarray}
\Gamma_{\mu}^{bsg}=-{G_F\over \sqrt{2}}  V_{ts}^*V_{tb}
{g_s\over 4\pi^2} (\Delta F_1 \bar s( q^2 \gamma_\mu - q\!\!\!/\  
q_\mu) LT^a b - i F_2 m_b \bar s \sigma_{\mu\nu}q^\nu RT^ab) .
\label{DECOM}
\end{eqnarray}
In the above, we have defined the form
factors $\Delta F_1$ and $F_2$ according to the convention in 
Ref. [4].  We have 
\begin{eqnarray}
&&\Delta F_1 ={4\pi \over \alpha_s} (C_4(\mu)+C_6(\mu)),\;\;
F_2 =-2C_8(\mu)
\label{F12}
\end{eqnarray}
We note that our relative signs of $\Delta F_1$ and $F_2$  agree with 
those in Refs. [4,6] and [17], and shall
result in a destructive interference~\footnote{We thank A. Kagan for
discussions which clarified this point.}.
At the NLL level, $\Delta F_1$ is corrected by $\Delta \bar{F}_1
\equiv {4\pi \over \alpha_s} (\bar{C}_4(q^2,\mu)+\bar{C}_6(q^2,\mu))$.

To obtain the branching ratio for $b\to s g \eta'$ from $b\to s g^*$ vertex, we 
use the anomalous $\eta'-g-g$ coupling given by: 
$a_g(\mu) \cos\theta \epsilon_{\mu\nu\alpha\beta}q^\alpha k^\beta$ with
$a_g(\mu) = \sqrt{N_F}
\alpha_s(\mu)/\pi f_{\eta'}$, q and k the momenta of the two gluons.

In previous one-loop calculations without QCD corrections, 
it was found that $\Delta F_1\approx -5$
and $F_2\approx 0.2$~\cite{AS,HT}.
In our approach, we obtain $\Delta F_1=-4.86$ and $\Delta F_2=0.288$ 
from Eqs. (\ref{WILSON}) and (\ref{F12}). However, $\Delta F_1$ is enhanced 
significantly by the matrix-element correction
$\Delta \bar{F}_1(q^2,\mu)$. The latter quantity develops 
an imaginary part as $q^2$ passes the charm-pair threshold, 
and the magnitude of its real part also
becomes maximal at this threshold. From Eqs. (3), (4) and (5), one  
finds 
Re$(\Delta \bar{F}_1(4m_c^2,\mu))=-2.58$ at $\mu=5$ GeV.
Including the contribution by $\Delta \bar{F}_1(q^2,\mu)$ with $\mu=5$ GeV, 
we find $B(b\to sg\eta')=5.6\times  
10^{-4}$
with a cut on $m_{X}\equiv \sqrt{(k+p')^2}\leq 2.35$ GeV. 
We also obtain the spectrum         
$dB(b\to sg\eta')/dm_X$ as depicted in Fig. 1. 
The peak of  
the 
spectrum corresponds to $m_X\approx 2.4$ GeV. 
The destructive interference of between $F_1$ and $F_2$ lowers down the 
branching ratio by about 14\% which is quite different from the results 
obtained in Refs.[3,4] because our $\Delta F_1$ is larger than
theirs.

In our calculation, $a_g(\mu)$ of the  
$\eta'-g-g$ vertex is treated as a constant independent 
of invariant-masses of the gluons, and $\mu$ is 
set to be $5$ GeV. In practice, $a_g(\mu)$ should behave like 
a form-factor which becomes suppressed as the gluons attached to it 
go farther off-shell~\cite{KP}. 
It is possible that the branching ratio 
we just obtained gets reduced significantly by the 
form-factor effect in $\eta'-g-g$ vertex.
Should a large form-factor suppression occur, the additional 
contribution from $b\to \eta' s$ and $B\to \eta' s \bar q$ 
discussed earlier will become crucial. 
We however like to stress that 
our estimate of $b\to sg\eta^{\prime}$
with $\alpha_s$ evaluated at $\mu=5$ GeV is conservative. To  
illustrate 
this, let us compare branching ratios for $b\to sg\eta^{\prime}$ 
obtained at $\mu=5$ GeV
and $\mu=2.5$ GeV respectively. 
The branching ratios at the 
above two scales with the kinematical cut on $m_X$ are $4.9\times 10^{-4}$ and 
$8.5\times 10^{-4}$ respectively. One can clearly see the significant
scale-dependence! With the enhancement resulting from lowering the 
renormalization scale, there seems to be some room for the  
form-factor 
suppression in the attempt of explaining $B\to \eta^{\prime}X_s$ by
$b\to sg\eta^{\prime}$. 
We do notice that
$B(b\to sg\eta^{\prime})$ is suppressed by more than one order of magnitude  
if $a_g(\mu)$ is replaced by $a_g(m_{\eta'})\cdot  
m_{\eta'}^2/
(m_{\eta'}^2-q^2)$ according to Ref.[6]. 
However, as pointed out in Ref.[4], the validity of such a prescription 
remains controversial.    

Before closing we would like to comment on the branching ratio for
$B\to \eta X_s$. It is interesting to note that the width of $b\to \eta s g$  
is suppressed by $\tan^2\theta$ compared to that of 
$b\to\eta' sg$.
Taking $\theta=-15.4^o$, we obtain  
$B(B\to \eta X_s)\approx 4\times 10^{-5}$.
The contribution from four-quark operator can be  
larger. 
Depending on the choice of parameters, we find that $B(B\to \eta X_s)$
is in the range of $(6\sim 10)\times 10^{-5}$.

In conclusion, we have calculated the branching ratio of 
$b\to sg\eta^{\prime}$ by including the NLL correction to the $b\to  
sg^*$
vertex. By assuming a low-energy $\eta^{\prime}-g-g$ vertex, we  
obtain
$B(b\to sg\eta^{\prime})=(5-9)\times 10^{-4}$ depending on the
choice of the QCD renormalization-scale. Although  
the form-factor suppression
in the $\eta^{\prime}-g-g$ vertex is anticipated, it remains possible  
that the anomaly-induced 
process $b\to sg\eta^{\prime}$ could account for the CLEO measurement
on the $B\to \eta^{\prime}X_s$ decay.  
For the four-quark operator contribution, we obtain
$B\to \eta' X_s \approx 1\times 10^{-4}$. This accounts for roughly  
15\% of the
experimental central-value and can reach 30\% if favourable  
parameters are
used. 

\begin{figure}[htb]
\centerline{\DESepsf(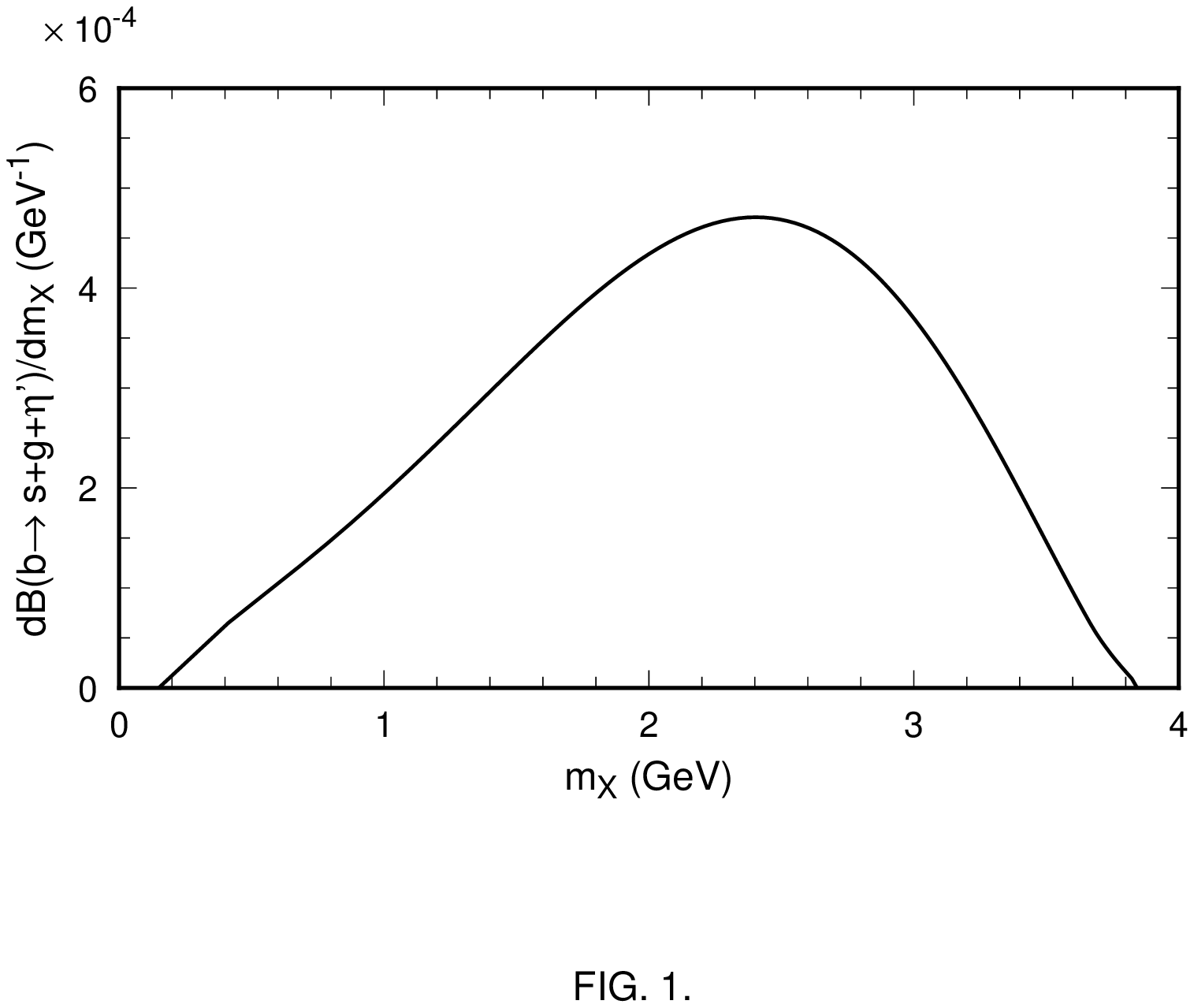 width 6 cm)} 
\caption{
The distribution of $B(b\to s+g+\eta')$
as a function of the recoil mass 
$m_X$.
}
\end{figure}
\section*{Acknowledgments} 
We thank W.-S. Hou, A. Kagan and A. Soni for discussions.
The work of XGH is supported by Australian Research 
Council and National Science Council of R.O.C. under the grant number
NSC 87-2811-M-002-046. The work of GLL is supported by
National Science Council of R.O.C. under the grant numbers 
NSC 87-2112-M-009-038, NSC 88-2112-M-009-002, 
and National Center for Theoretcal Sciences of  
R.O.C. 
under the topical
program: PQCD, B and CP.

\end{document}